\documentclass[hidelinks,11pt]{article}

\usepackage{fullpage}
\usepackage{amsmath, amssymb}
\PassOptionsToPackage{hyphens}{url}\usepackage{hyperref}
\usepackage{graphicx}
\usepackage[singlelinecheck=false, labelsep=space, labelfont=bf]{caption}
\usepackage[labelformat=simple]{subcaption}
\usepackage{tabularx}
\usepackage{multirow}
\usepackage[round]{natbib}
\usepackage{arydshln}

\usepackage[notablist, nomarkers, tablesfirst]{endfloat}
\usepackage{setspace}
\DeclareCaptionSubType[roman]{table}
\title{\vspace{-2.0cm}Home Sweet Home: Quantifying Home Court Advantages For NCAA Basketball Statistics}
\author{Matthew van Bommel$^1$\thanks{Corresponding author: Matthew van Bommel, Department of Statistics and Actuarial Science, Simon Fraser University, 8888 University Drive, Burnaby, B.C., Canada, V5A 1S6. E-mail: matthew.vbommel@gmail.com} \\  \and Luke Bornn$^1$ \and Peter Chow-White$^{2,3}$ \and Chuancong Gao$^3$}
\date{\small{\textit{$^1$ Department of Statistics and Actuarial Science, Simon Fraser University, Burnaby, Canada \\
$^2$ School of Communication, Simon Fraser University, Burnaby, Canada\\
$^3$ School of Computing Science, Simon Fraser University, Burnaby, Canada}}}

\begin{document}
	
	\maketitle
	
	\begin{abstract}
		Box score statistics are the baseline measures of performance for National Collegiate Athletic Association (NCAA) basketball. Between the 2011-2012 and 2015-2016 seasons, NCAA teams performed better at home compared to on the road in nearly all box score statistics across both genders and all three divisions. Using box score data from over 100,000 games spanning the three divisions for both women and men, we examine the factors underlying this discrepancy. The prevalence of neutral location games in the NCAA provides an additional angle through which to examine the gaps in box score statistic performance, which we believe has been underutilized in existing literature. We also estimate a regression model to quantify the home court advantages for box score statistics after controlling for other factors such as number of possessions, and team strength. Additionally, we examine the biases of scorekeepers and referees. We present evidence that scorekeepers tend to have greater home team biases when observing men compared to women, higher divisions compared to lower divisions, and stronger teams compared to weaker teams. Finally, we present statistically significant results indicating referee decisions are impacted by attendance, with larger crowds resulting in greater bias in favor of the home team. 
	\end{abstract}

\clearpage

\section{Introduction}
\label{introduction}
The home team advantage is an easily observed phenomenon in sports, particularly in National Collegiate Athletic Association (NCAA) basketball. At the end of the 2017-2018 season, 337 of 351 Division I men’s basketball teams had a historical winning record at home, with a median home winning percentage of 67.7\% \citep{rpi_ratings}. However, it is not only the win-loss record that is affected by playing at home; team and player box score statistics also differ between home and away teams. While many previous studies have examined home advantages in sports, few have examined the impact on box score statistics, and those that have tend to focus on professional leagues \citep{park_factors,nhl_rink_factors,scorekeeper_bias}. In this paper, we will conduct an examination of both the neutral and home court impacts on box score statistics in NCAA basketball, across both genders and all three divisions.

Box score statistics lie at the heart of evaluating the performance of NCAA basketball players and teams. Coaches and players frequently use these statistics to gain a better understanding of the playing style and ability of opposing teams and players, particularly in preparing for upcoming games. Statistics are also influential in the process of selecting end of season awards at both the player and team level. Additionally, professional basketball leagues, including the National Basketball Association (NBA) and the Women’s National Basketball Association (WNBA), use box score statistics in their evaluation of college players when deciding on draft selections or free agent signings. In this way, box score statistics can impact the success of professional franchises, not to mention the lives of the players themselves. These impacts outline the importance of understanding the process behind NCAA box score statistics and any influential factors. 

Traditional views of home court advantage support the impact of travel, spectator support, and home team familiarity \citep{measurement_interpretation_home_advantage}. More recent attention has been focused on the impact of human biases through participants such as referees \citep{new_insights_home_team_advantage} and scorekeepers \citep{scorekeeper_bias}. Specific to the NCAA, previous research has concluded basketball referees call more fouls on away teams \citep{ncaa_officiating_bias}. Additionally, the teams of scorekeepers (or statisticians as they are sometimes referred) tracking box score statistics are not a neutral party but are employed by the home teams of NCAA games. Given that the NCAA itself admits “at times, statisticians have to use their judgment and knowledge on how to score a certain play” \citep{huntsville_assists}, the possibility of scorekeeper bias in favor of the home team, either intentional or otherwise, seems a relevant concern. Thus, this report will examine the human bias factors for referees and scorekeepers in greater detail.

NCAA data has unique challenges compared to data from professional leagues. Even when restricting to a single gender-division combination, NCAA basketball contains far more teams playing fewer games over the course of a season than its professional counterparts the NBA and WNBA. This reality makes it difficult to draw conclusions at the individual team level. However, NCAA data also has its advantages. A feature of NCAA schedules is that teams play games each season at neutral locations: a location not considered to be the home of either team in the game. Such games provide a unique tool in examining home court advantage that has been underutilized in the existing literature. While occasionally used in quantifying an overall home court advantage \citep{home_court_advantage}, to our knowledge neutral location data has never been used in quantifying home court impact on box score statistics. Additionally, having multiple genders and divisions under the same organizational structure allows us to more easily examine factors such as player gender, team skill, and attendance. This report aims to utilize these advantages in presenting a full picture of the home court advantage on box score statistics in NCAA basketball.  

The remainder of this paper is organized as follows. Section 2 outlines the data used in the paper and the statistics that will be examined. An overall view of home court advantage is presented using summary statistics in Section 3, and a selection of interesting observations are discussed in greater detail. Sections 4 and 5 present the results of statistical models that dive deeper into the home court advantages. The former section examines the impact of attendance on box score statistics and the latter quantifies the relative impact of home court and other factors on statistic totals. Finally, Section 6 ends with a summary of conclusions and a discussion.

\section{Data}
\label{data}
To obtain the data used in this report, we scraped box score information from stats.ncaa.org for 153,570 games between the 2011-2012 and the 2015-2016 seasons across Division I (D1), Division II (D2), and Division III (D3) NCAA basketball for both women and men. There were 11,930 games with missing or inconsistent data and these games were removed from the dataset before performing any of the analyses presented in this report. Of the removed games, 14 failed our basketball focused data quality checks which included verifying that at least 1 field goal attempt and rebound were recorded by each team and that the number of assists for a team was less than or equal to the number of made field goals. The remainder of the removed games were missing data for one or more box score statistics. In either case the full game, rather than simply the affected statistics, were removed from consideration. The gender-division combinations were affected relatively equally by these data quality issues, with the proportion of data included in the final dataset ranging from 90\% (D3 women) to 94\% (D1 men). At the team level, there are a small number of teams with most of their data missing, however 96\% of teams have at least 80\% of their available games included in the dataset. Overall, our final dataset contains 141,640 games with full information for the box score statistics examined in this report. The full list of examined statistics, along with their abbreviations, is presented in Table \ref{box_score_abbreviations}.

In addition to standard box score information, we also obtain the attendance and location of the games, including whether each team played at home, away, or at a neutral location. In our final dataset, 5,847 games were missing attendance information and thus were removed to conduct the attendance focused analyses presented in Section \ref{the_crowd_effect}. The home-away focused analyses of Section \ref{quantifying_home_court_advantage} were conducted after removing the 23,856 neutral location games. Such games, in which neither team is playing at home, are comprised of games played in mid-season invitational tournaments and end of season NCAA tournaments and make up between 7\% and 11\% of the data for each gender-division combination. These games tend to have slightly stronger teams participating (measured by rating percentage index) and slightly lower attendance, which the exception of D1 neutral location games, which tend to have notably higher attendance.

\begin{table}[!t]
	\centering
	\caption{Box score statistics and their abbreviations}
	\label{box_score_abbreviations}
	\begin{tabular}{ l l }
		\hline\noalign{\smallskip}
		\textbf{Abbreviation} & \textbf{Statistic} \\
		\noalign{\smallskip}\hline\noalign{\smallskip}
		3FGA & 3 point field goal attempts \\
		3FG\% & 3 point field goal percentage \\
		AST & Assists \\
		BLK & Blocks \\
		DREB & Defensive rebounds \\
		FGA & Field goal attempts \\ 
		FG\% & Field goal percentage \\
		FTA & Free throw attempts \\
		FT\% & Free throw percentage \\
		OREB & Offensive rebounds \\
		PF & Personal fouls \\
		PTS & Points \\
		STL & Steals \\
		TOV & Turnovers \\
		\hline\noalign{\smallskip}
	\end{tabular}
\end{table}

\section{Home Court Advantage}
\label{home_court_advantage}
In all seasons between 2011-2012 and 2015-2016, NCAA basketball teams performed better at home compared to on the road in nearly all statistical categories across both genders and all three divisions. The magnitudes of these performance advantages for all statistic-gender-division combinations are displayed in Figure 1. The average overall home boosts for statistics ranged from 0.9\% (FGA) to 18.8\% (BLK), with values at the season level reaching all the way to 30.7\% for BLK in 2014 men's D1. The percent increase value for a given combination is computed by filtering the data to include only games for the gender-division of interest, subtracting the average away team statistic value from the average home team statistic value, and dividing the difference by the overall average statistic value in the filtered dataset. The average relative variance of the distribution of statistic home advantages across seasons is 0.15, compared to 0.80 for statistic distributions across gender-division combinations. Thus, within a combination, home advantages are relatively stable across seasons, implying consistent underlying factors (and not randomness) as the cause. We can further examine home-away differences by splitting them into home-neutral differences and neutral-away differences (computed using the same logic as above), the distributions for which are displayed in Figure 2. Note that summing the home-neutral and neutral-away values for each statistic would reproduce the results of Figure 1. 

\begin{figure*}
	\centering
	\includegraphics[width=0.9\textwidth]{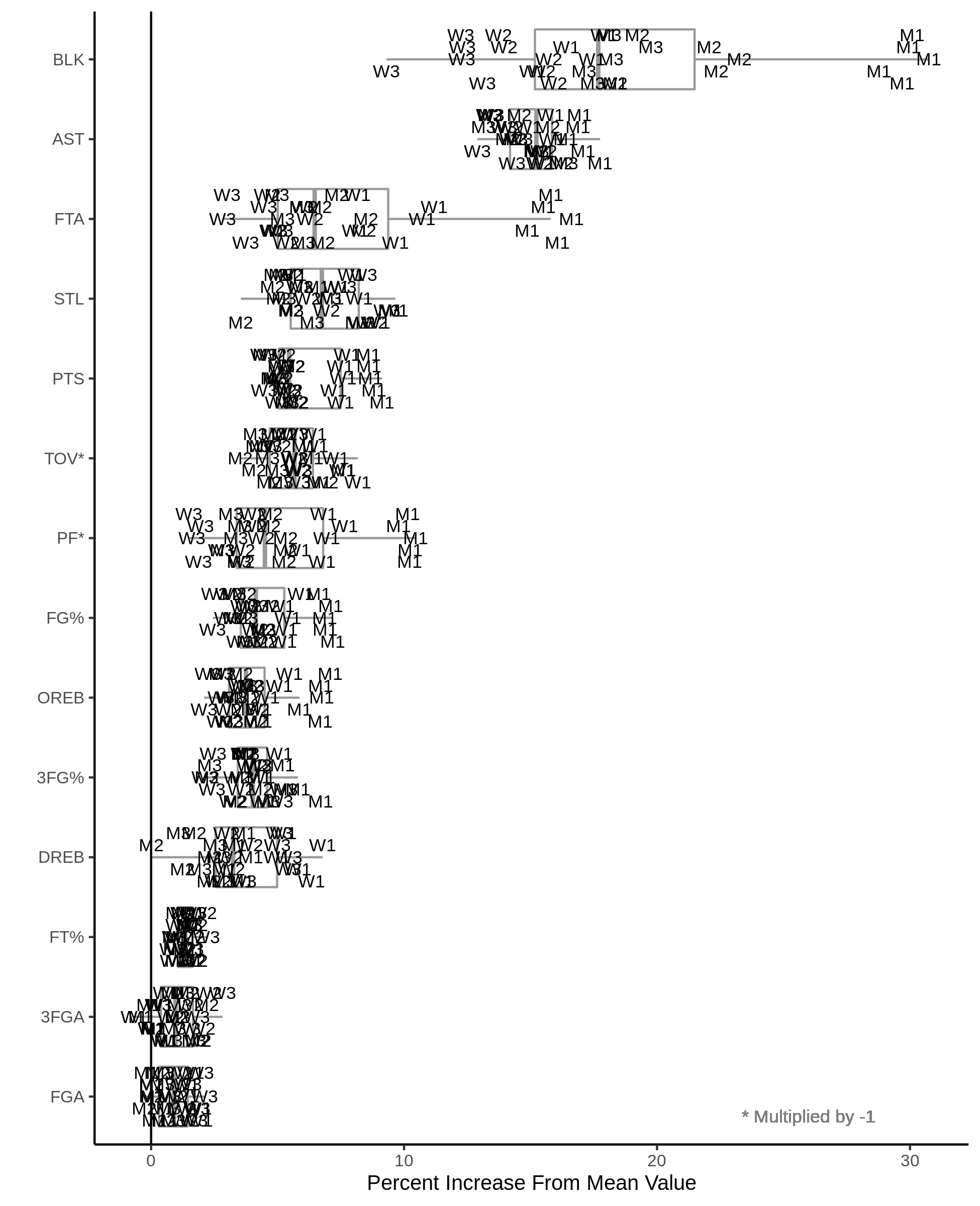}
	\caption{Distributions of the percent increase for home teams compared to away teams for a variety of box score statistics, across all gender-division-season combinations, sorted by the mean percent increase. The individual observations represent a specific combination of gender (labeled letter), division (labeled number), and season (ordered vertically with 2011-2012 at the top and increasing down until 2015-2016 at the bottom). Note that positive values indicate an improvement for each statistic (decrease in PF and TOV and increase in all other statistics).}
	\label{home_away_box_score_comparison}
\end{figure*}

\begin{figure*}
	\centering
	\includegraphics[width=0.9\textwidth]{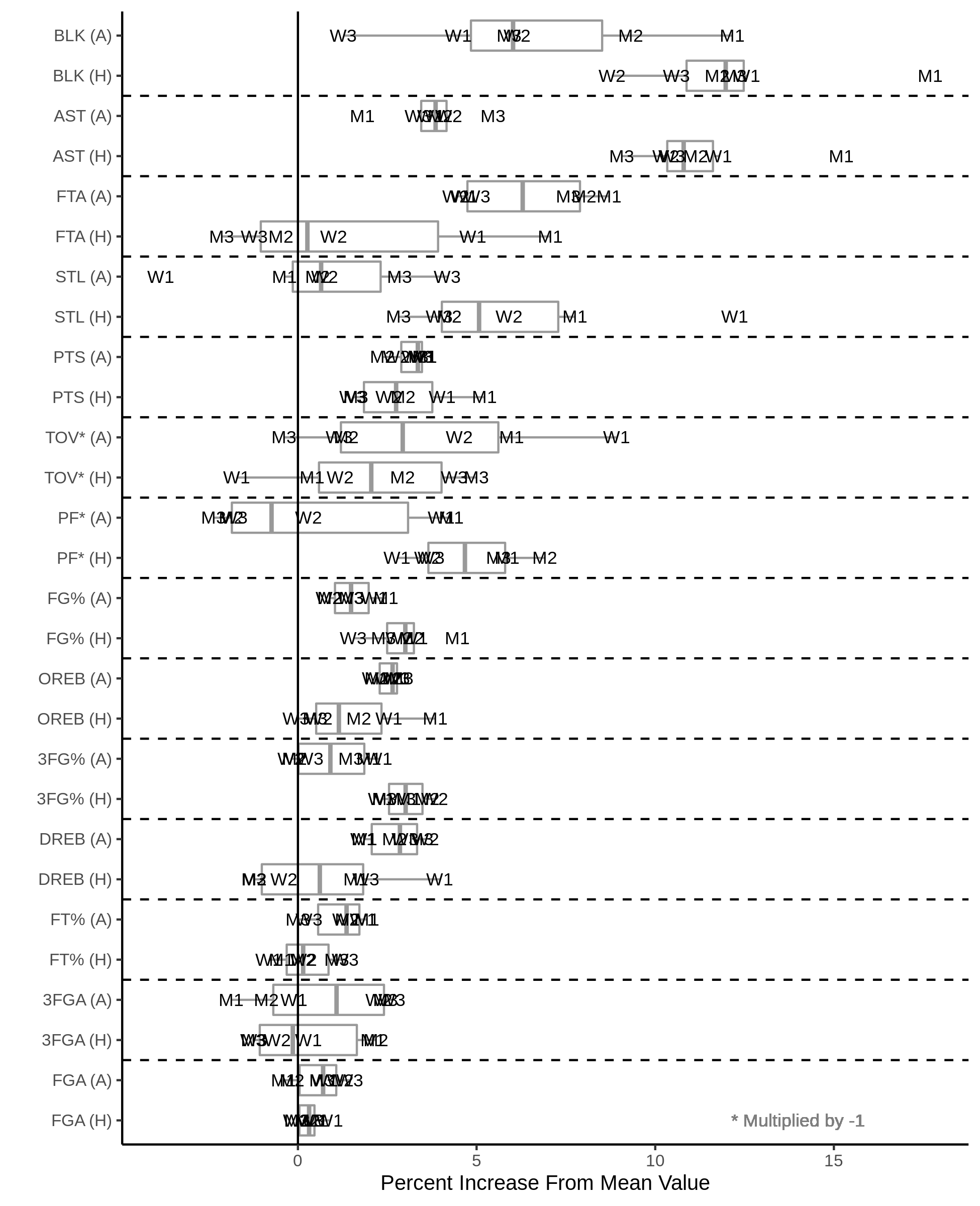}
	\caption{Distributions of the percent increase for home teams compared to neutral teams (H) and for neutral teams compared to away teams (A) for a variety of box score statistics, across all gender-division combinations. The individual observations represent a specific combination of gender (labeled letter) and division (labeled number) across all seasons. Note that positive values indicate an improvement for each statistic (decrease in PF and TOV and increase in all other statistics) in the expected direction (home team improvement for home-neutral and neutral team improvement for neutral-away).}
	\label{home_away_neutral_box_score_difference_distributions}
\end{figure*}

The combination of these figures presents many interesting insights. Here we list several such insights, and the subsections below dive deeper into two deserving a closer look:

\begin{enumerate}
\item AST and BLK, largely considered the most subjective statistics, have noticeably greater home advantages compared to the other more objective statistics
\item the effect magnitudes for nearly all statistics examined are noticeably impacted by division, and many statistics are also impacted by gender
\item the average home - away advantages in 3FG\% and FG\% are driven by the difference between home and neutral team performance (accounting for 76\% and 65\% of the advantages respectively), while neutral teams experience a relatively minor boost compared to away teams
\item FT\% is negatively impacted for away teams compared to neutral teams (-1.1\%), but home teams barely receive an additional boost compared to neutral teams (0.2\%) and neutral D1 teams actually have a higher FT\% than home teams
\item as would be expected given the relationship between the two statistics, the average home advantage of PF (4.8\%) is a similar magnitude to the away disadvantage of FTA (-6.4\%)
\end{enumerate}

\subsection{The Impact of Scorekeeper Subjectivity}
\label{impact_of_scorekeeper_subjectivity}
There are likely several factors contributing to the home advantage observed for box score statistics, including superior on-court performance. However, it is unlikely that such a performance improvement would result in AST (15.1\%) and BLK (18.8\%) having noticeably greater average home advantages compared to the other statistics (0.9\% to 7.9\%). According to a former NBA scorekeeper, scorekeepers are “given broad discretion over two categories: assists and blocks” \citep{confessions_of_nba_scorekeeper}. This statement aligns with previous observations on the subjectivity of statistics at both the NBA \citep{nba_misleading_number,nba_court_factors} and NCAA \citep{huntsville_assists} levels, including model-quantified patterns of inconsistent scorekeeper behavior for these statistics in the NBA \citep{scorekeeper_bias}. Additionally, the statistics with the least subjectivity (particularly FGA, 3FGA, and FT\%) are all among the statistics with the least advantage for the home team. Thus, these results support the hypothesis that scorekeeper biases are a major factor in the home court advantages for box score statistics.

From Figure 2 we also see that most of the advantage for AST (11.3\% vs 3.8\%) and BLK (12.3\% vs 6.5\%) comes from being the home team, and there is less of a difference between neutral and away teams. This result supports the idea that scorekeepers are more biased in favor of the home team than they are biased against the away team. The trend reverses for the more objective statistics such as FGA (0.3\% vs 0.6\%), 3FGA (0.3\% vs 0.8\%), and FT\% (0.2\% vs 1.1\%) with the boost provided to neutral teams over away teams being greater than that for home teams over neutral teams. This reversal implies an increase in the relative impact of on-court performance for less subjective statistics.

\subsection{The Impact of Gender and Division}
\label{impact_of_gender_and_division}
Table 2 presents the statistics for which the order of the percent increase values (from Figures 1 and 2) is dictated by gender or division. Even when restricting to perfect sorting, division orders at least one of the home-away, home-neutral, or neutral-away distributions for 11 of the 14 statistics, including the 8 statistics with the greatest home-away differences. Only 4 of these statistics (STL, TOV, FT\%, and FGA) have a distribution ordered such that D3 teams receive the greatest boost and D1 the least (the rest have the reverse order). Of those 4 statistics, only STL and FGA have consistent ordering across distributions and none of the statistics see D3 have the greatest boost for the overall home-away distribution. We also observe similar trends for gender: 10 of the 14 statistics have at least one distribution sorted by gender, though only 4 have gender as a primary sort for a distribution. Of the 10 total statistics, only 3 (STL, TOV, and FGA) have a distribution for which women receive the greater advantage and only 1 (FGA) has women receive the greater boost for its home-away distribution. Overall, men and higher divisions tend to see greater impact in the expected directions (home over neutral over away) and this observation is especially true when comparing box score statistics between home teams and away teams.

\begin{table}[!t]
	\renewcommand{\arraystretch}{1}
	\centering
	\caption{Division and gender ordering of statistic home percent increase effects}
	\label{gender_division_results}
	\begin{tabular}{ l c c c }
		\hline\noalign{\smallskip}
		\textbf{} & \textbf{Home - Away} & \textbf{Home - Neutral} & \textbf{Neutral - Away} \\
		\noalign{\smallskip}\hline\noalign{\smallskip}
		\multirow{2}{*}{BLK} & M $>$ W & D1 $>$ D2 $>$ D3 & \\ &  &  & \\
		\hdashline
		\multirow{2}{*}{AST} & D1 $>$ D2 $>$ D3 &  &  \\ & M $>$ W &  & \\
		\hdashline
		\multirow{2}{*}{FTA} & D1 $>$ D2 $>$ D3 & D1 $>$ D2 $>$ D3 & M $>$ W \\ & M $>$ W &  & \\
		\hdashline
		\multirow{2}{*}{STL} &  & D3 $>$ D2 $>$ D1 & D3 $>$ D2 $>$ D1 \\ &  & W $>$ M & \\
		\hdashline
		\multirow{2}{*}{PTS} & D1 $>$ D2 $>$ D3 & D1 $>$ D2 $>$ D3 & \\ &  & M $>$ W & \\
		\hdashline
		\multirow{2}{*}{TOV} &  & D3 $>$ D2 $>$ D1 & D1 $>$ D2 $>$ D3 \\ &  & M $>$ W & W $>$ M \\
		\hdashline
		\multirow{2}{*}{PF} & D1 $>$ D2 $>$ D3 & M $>$ W & \\ & M $>$ W &  & \\
		\hdashline
		\multirow{2}{*}{FG\%} &  & D1 $>$ D2 $>$ D3 &  \\ &  & M $>$ W & \\
		\hdashline
		\multirow{2}{*}{3FG\%} &  &  &  \\ &  &  & \\
		\hdashline
		\multirow{2}{*}{OREB} &  & D1 $>$ D2 $>$ D3 &  \\ &  & M $>$ W & \\
		\hdashline
		\multirow{2}{*}{DREB} &  &  &  \\ &  &  & \\
		\hdashline
		\multirow{2}{*}{FT\%} &  & D3 $>$ D2 $>$ D1 & D1 $>$ D2 $>$ D3 \\ &  &  & \\
		\hdashline
		\multirow{2}{*}{3FGA} &  &  &  \\ &  &  & \\
		\hdashline
		\multirow{2}{*}{FGA} & W $>$ M &  & W $>$ M \\ &  &  & D3 $>$ D2 $>$ D1 \\
		\hline\noalign{\smallskip}
		\multicolumn{4}{p{11cm}}{\raggedright The top lines indicate the primary ordering and the bottom lines indicate a secondary ordering. The statistics are sorted to allow easy comparability with Figures \ref{home_away_box_score_comparison} and \ref{home_away_neutral_box_score_difference_distributions}.}
	\end{tabular}
\end{table}

Interestingly, of the 3 statistics for which gender and division impact all the overall home-away distributions, one (AST) is arguably the most subjective statistic recorded by the scorekeepers and the other two (FTA and PF) are the referee driven statistics. Specifically, the home-away impacts on both FTA and PF are both sorted by division, with D1 teams receiving a noticeably greater boost. Knowing that average attendance values are also ordered according to division, we raise the hypothesis that referee bias is impacted by attendance. Such a hypothesis aligns with previous studies claiming home win probability advantages are primarily due to referee bias \citep{scorecasting}, and that soccer referee bias in favor of the home team increases as attendance increases \citep{favoritism_under_social_pressure}. The following section examines this hypothesis in greater detail, by examining the impact of attendance on all box score statistics.

\section{The Crowd Effect: Attendance Impact on Statistics}
\label{the_crowd_effect}
This section seeks to determine the importance of fans in a game, specifically by answering the question: do larger crowds have a positive impact on box score statistics for the home team? To test the impact of attendance on box score statistics for a gender-division combination, we compare the home advantages (home team statistic value – away team statistic value) in high attendance games and low attendance games for a variety of box score statistics. Specifically, we perform two-sample t-tests for significant differences in the mean home advantage values. Here, we define low and high attendance games as games with attendance values in the bottom 25 percent and top 25 percent respectively of the corresponding gender-division attendance values. The cutoffs for each gender-division combination are presented in Table 3.

From the data, we observe an intuitive trend which complicates our test procedure: home teams with higher attendance tend to be stronger teams. This difference in strength must be controlled in order to to draw meaningful conclusions on the impact of attendance. One method of measuring team strength is rating percentage index (RPI). More sophisticated measures of team strength have been developed but we choose RPI for its interpretability and relevance in NCAA history \citep{wikipedia_rpi}.  We calculate the RPI for a team in a given season as:

\[
RPI = \left(WP \times 0.25 \right) + \left(OWP \times 0.50 \right) + \left(OOWP \times 0.25 \right)
\]

\noindent where WP is team winning percentage, OWP is opponents' winning percentage, and OOWP is
opponents' opponents' winning percentage. For a measure of relative team strength in a game, we
compute the home team RPI advantage ($\textrm{RPI}'$) as:

\[
RPI' = Home \ Team \ RPI - Away \ Team \ RPI
\]

Across all gender-division combinations, the two sided Kolmogorov–Smirnov (KS) test, performed using the R function \texttt{ks.test} in the \texttt{stats} package \citep{R}, concludes the distributions of $\textrm{RPI}'$ in high attendance games are significantly different than those for low attendance games. A visual example of the $\textrm{RPI}'$ difference for D1 women is presented in Figure 3. Thus, any potential difference in home advantages between high attendance and low attendance games caused by attendance, will be confounded by $\textrm{RPI}'$ differences. 

\begin{figure*}
	\centering
	\includegraphics[width=\textwidth]{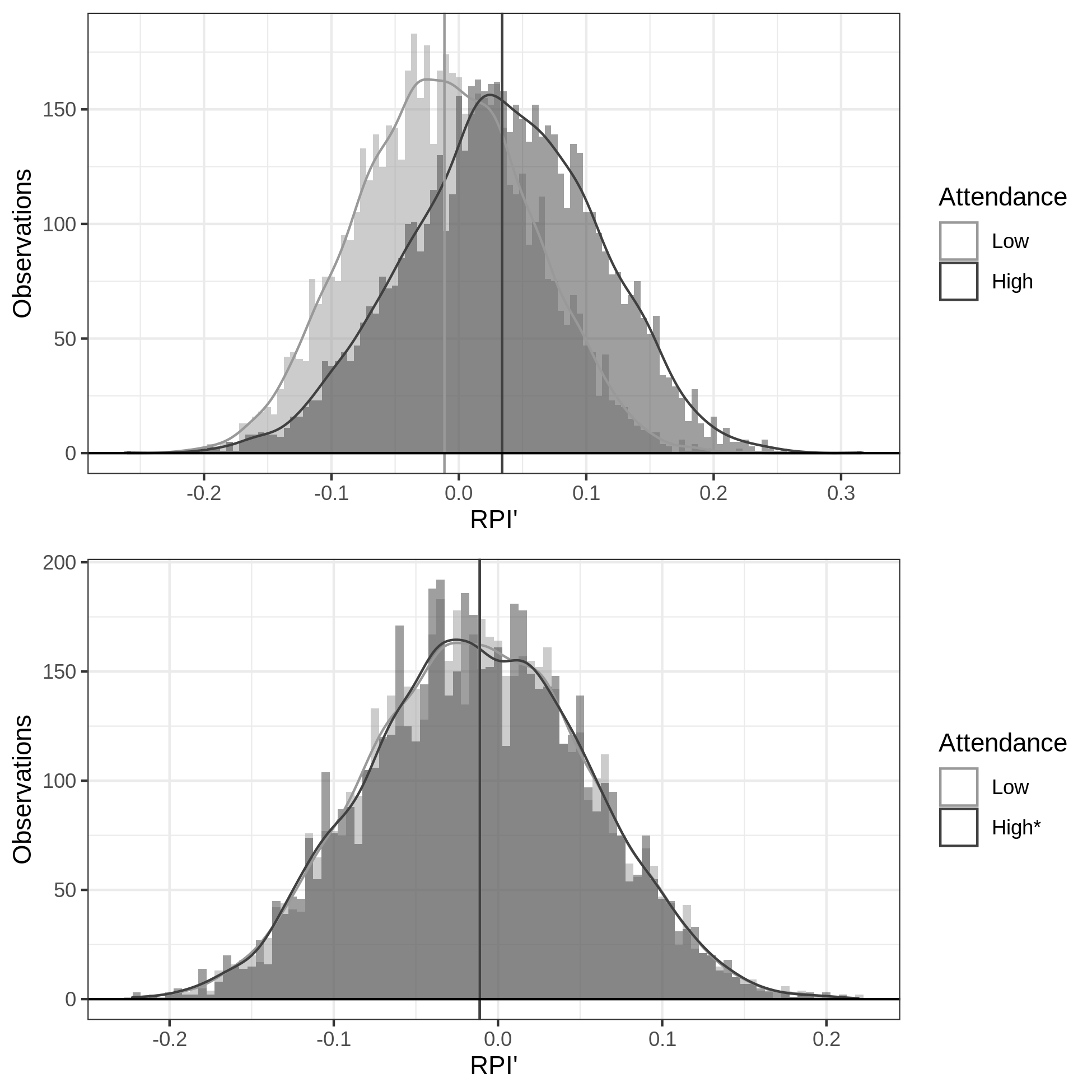}
	\caption{Results of $\textrm{RPI}'$ statistical matching for 2015-2016 D1 women data. The Low and High attendance distributions are those observed in the data and the High* distribution is the result of the matching algorithm. The vertical lines label the mean of each distribution.}
	\label{rpi_matching}
\end{figure*}

To remove the $\textrm{RPI}'$ impact on the home advantage results, we use statistical matching techniques to transform our high attendance samples so that their $\textrm{RPI}'$ distributions match the corresponding low attendance $\textrm{RPI}'$ distributions. Statistical matching provides an intuitive approach to controlling for $\textrm{RPI}'$, free of any assumptions on the form of its relationship to the box score statistics. This feature allows a common technique to be applied to all test procedures, even though the relationships with $\textrm{RPI}'$ differ among the individual statistics. Specifically, given distributions of low attendance and high attendance games, we perform the following algorithm:
\begin{itemize}
\item  Define 25 equal width bins $b_1, \ldots, b_{25}$ across the range of $\textrm{RPI}'$ for the low attendance distribution
\item  For each bin $b_i$:
	\begin{itemize}
		\item Determine the number of low attendance observations with $\textrm{RPI}'$ values in $b_i$ and define this value as $n_i$
		\item Randomly sample with replacement from the high attendance observations with $\textrm{RPI}'$ values within $b_i$ until $n_i$ values have been sampled and define this set of values as $H_i$. If there are no high attendance observations with $\textrm{RPI}'$ values within $b_i$, define $H_i$ to be the empty set.
	\end{itemize}
\item  Combine the values of $H_i, i = 1, \ldots, 25$ to form a new distribution of high attendance game observations, replacing the original distribution of high attendance game observations
\end{itemize}

Sample results of the above algorithm for 2015-2016 D1 women are presented in Figure 3. Matching visual intuition, the KS test concludes there are no significant differences in distribution between the low attendance observations and the adjusted high attendance observations resulting from the above algorithm, across all gender-division combinations.

With the impact of $\textrm{RPI}'$ mitigated, we are able to perform two-sample t-tests to test for significant differences in the mean home advantage values, using the R function \texttt{t.test} in the \texttt{stats} package \citep{R}. Since the matching algorithm has a random component, we repeat the process 1000 times for each statistic, fitting 1000 distinct models. The average differences in mean home advantage values between the high and low attendance samples are displayed in Table 3, with negative values implying an advantage for the away teams. The bold values in the table are those with significant differences between the samples. Since we examine 14 statistics and 6 gender-division combinations, we perform 84 tests and thus use the Bonferroni correction to select a level of significance of $\alpha=0.05/84=0.0006$. The two statistics which have significantly different home advantage means in the low attendance and high attendance subsets for the most division-gender combinations are PF and FTA: the two referee driven statistics. This result supports our hypothesis from Section 3, providing strong evidence that referee bias in favor of the home team increases as attendance increases. 

\begin{table}[!t]
	\centering
	\caption{Results of two-sample t-tests for significant differences in the mean home advantage values between low attendance and high attendance games}
	\label{two_sample_t_test_results}
	\small
	\begin{tabular}{ l r r r r r r }
		\hline\noalign{\smallskip}
		\textbf{} & \multicolumn{2}{c}{\textbf{Division I}} & \multicolumn{2}{c}{\textbf{Division II}} & \multicolumn{2}{c}{\textbf{Division III}} \\
		\textbf{} & \multicolumn{1}{c}{\textbf{Men}} & \multicolumn{1}{c}{\textbf{Women}} & \multicolumn{1}{c}{\textbf{Men}} & \multicolumn{1}{c}{\textbf{Women}} & \multicolumn{1}{c}{\textbf{Men}} & \multicolumn{1}{c}{\textbf{Women}} \\ 
		\hline\noalign{\smallskip}
		Low  & 1500  & 426   & 283   & 180   & 181   & 112   \\
		High & 6943  & 1800  & 901   & 526   & 500   & 300   \\
		\hline\noalign{\smallskip}
		PF              & \textbf{0.87}  & \textbf{0.75}  & \textbf{0.55}  & 0.41  & \textbf{0.61}  & \textbf{0.92}  \\
		FTA             & \textbf{1.38}  & \textbf{0.92}  & \textbf{1.03}  & 0.59  & 0.77  & \textbf{1.21}  \\
		FGA             & \textbf{-0.87} & -0.43 & -0.03 & -0.03 & -0.28 & -0.44 \\
		BLK             & \textbf{0.27} & -0.01 & 0.25  & 0.10  & 0.17  & -0.03 \\
		TOV             & 0.18  & -0.01 & 0.21  & 0.04  & 0.31  & 0.32  \\
		PTS             & 0.36  & 0.38  & \textbf{1.61}  & 1.13  & \textbf{1.15}  & 0.90  \\
		AST             & -0.13 & 0.16  & 0.27  & 0.43  & 0.43  & -0.01 \\
		FG\%            & 0.29  & 0.04  & \textbf{0.97}  & 0.57  & 0.59  & 0.11  \\
		DREB            & -0.37 & 0.00  & 0.31  & 0.26  & -0.17 & -0.14 \\
		3FGA            & -0.58 & 0.34  & \textbf{-0.77} & -0.21 & -0.02 & 0.14  \\
		OREB            & 0.06  & 0.08  & \textbf{0.69}  & 0.50  & 0.21  & 0.03  \\
		STL             & 0.01  & -0.11 & 0.14  & -0.03 & 0.19  & 0.04  \\
		FT\%            & 0.32  & -0.33 & 0.07  & 0.32  & 0.19  & 0.52  \\
		3FG\%           & 0.15  & -0.25 & 0.34  & 0.43  & 0.33  & 0.29  \\
		\hline\noalign{\smallskip}
		\multicolumn{7}{p{10cm}}{\raggedright The Low and High rows present the cutoffs for low and high attendance games respectively for each gender-division combination. The remaining rows present sample mean differences averaged over 1000 iterations of t-tests, with bold values indicating those with average p-values significant at the level $\alpha=0.0006$, selected using the Bonferroni correction.}
	\end{tabular}
\end{table}
\normalsize

The only other statistic with significant differences for multiple gender-division combinations is PTS. Since PTS are also influenced by referees (through FTA), this result is likely a consequence of the home advantage in PF and FTA. Additionally, the observed PTS differences for D2 and D3 are noticeably greater than those for D1, implying that referee decisions may have greater impact on outcomes at levels in which the teams and players are less skilled. Interestingly, FT\% is not significantly impacted by attendance in any gender-division combination. Thus, while playing at home has a positive effect on FT\% (see Figure 1), the size of the home crowd does not seem to have a significant impact.

\section{Quantifying Home Court Advantage}
\label{quantifying_home_court_advantage}
We now seek to quantify the impact of playing at home on the totals of team box score statistics. In constructing the models for each statistic, we use data from the 2015-2016 season after removing the neutral location games from the dataset to directly target home-away differences. Additionally, given that we are constructing models for all games rather than comparing two subsets, we do not apply the matching procedure from the previous section. Since we are estimating count data, Poisson regression models seem a logical choice, however, the dispersion parameters across the 11 statistics range from 0.68 (FGA) to 3.01 (FTA). Additionally, using the R function \texttt{dispersiontest} in the \texttt{AER} package \citep{AER}, we reject the null hypothesis of equidispersion for all statistics expect DREB. Therefore, to keep the model form consistent across all statistics, we select the generalized Poisson regression model for its ability to control for both over and under dispersion \citep{generalized_poisson}. In particular, we use the R function \texttt{glm} with the \texttt{quasipoisson} family in the \texttt{stats} package \citep{R}.

\begin{figure*}
	\centering
	\includegraphics[width=\textwidth]{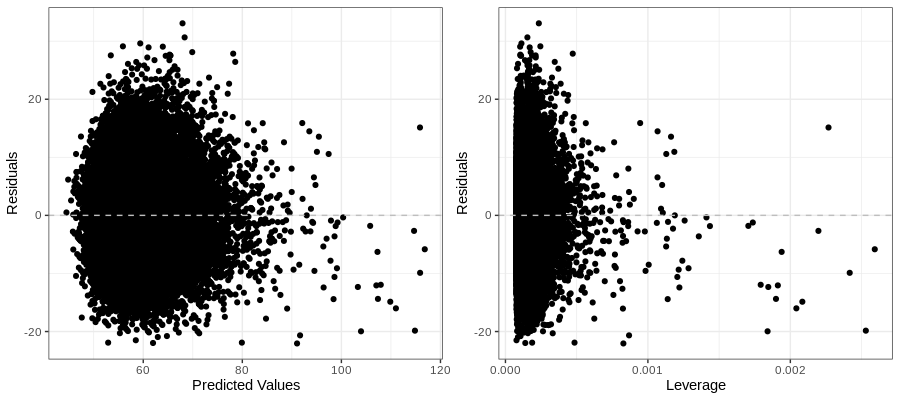}
	\caption{Residuals vs predicted values plot and residuals vs leverage plot for the FGA generalized Poisson regression model for 2015-2016 data.}
	\label{fga_residauls}
\end{figure*}

In addition to the home-away status, the models also control for number of possessions in the game, $\textrm{RPI}'$, and attendance (grouped into low, medium, and high as defined in Table \ref{two_sample_t_test_results}). These parameters are intuitively connected to statistic totals and we verify their importance by comparing deviance values for models with and without each parameter. Additionally, we observe significance for all parameters across the models for all statistics at the Bonferroni adjusted $\alpha=0.05/11=0.0045$. We also perform several diagnostic checks for each model. As an example, Figure \ref{fga_residauls} displays the residuals vs predicted values plot and the residuals vs leverage plot for the FGA model. The high leverage observations tend to also have high predicted values and negative residuals. Upon examination, these observations correspond to games in the top 0.1\% of observed possessions. Thus, high possession games are influential in the model and tend to be underestimated, a pattern we see repeated in the other models. However, for the FGA model shown and for the other models, this pattern affects only a small number of observations, and in general the residuals are 0 centered and equally distributed across the range of predicted values. Thus, we are confident in the reliability of our models to produce meaningful results. 

Having combined all gender-division combinations in building and verifying model performance, we now estimate unique models for each gender-division-statistic combination to examine the impacts within each combination. After estimating the models, we can convert the model coefficients, $\beta_j$, to percent impact values (the percent a statistic total is increased or decreased by the term corresponding to the coefficient) through the transformation $e^{\beta_j} - 1$. Percent impact values for home advantage are presented in Table \ref{percent_impact_values}. The table presents results for the 2015-2016 season, but testing other seasons produces similar conclusions. Even when controlling for the other factors in the model, the models select positive coefficients for home court impact for all statistics in all gender-division combinations, with only the exception of FGA in two of the six combinations. Further, 56 of the 66 total models produce fully positive 95\% confidence intervals for the home court coefficient, 9 models (all for FGA, 3FGA, or OREB) produce confidence intervals containing 0, and only 1 model (men's D1 FGA) produces a fully negative confidence interval for the home court coefficient.

Aligning with previous observations, BLK and AST have a much greater home team advantage compared to the other statistics, with home teams receiving over a 11\% boost for each gender-division combination and advantages stretching all the way to 25.4\% for men's D1. Also in alignment with results from previous sections, the models indicate men get greater average home court boosts than women for BLK (+9.06\%), AST (+1.08\%), and FTA (+1.59\%) - the three most impacted statistics. Additionally, PF and FTA remain among the statistics most impacted by home team advantage, even when controlling for attendance in the model. Thus, though these statistics were shown to be significantly impacted by attendance in the previous section, these results indicate there is an underlying home court impact, independent of attendance.

\begin{table}[!t]
	\centering
	\caption{Percent impact of home court advantage on the total count of box score statistics observed in games for the 2015-2016 NCAA season}
	\label{percent_impact_values}
	\begin{tabular}{ l r r r r r r r r r r r}
		\hline\noalign{\smallskip}
		\textbf{} & \textbf{BLK} & \textbf{AST} & \textbf{FTA} & \textbf{STL} & \textbf{PF} & \textbf{PTS} & \textbf{TOV} & \textbf{DREB} & \textbf{OREB} & \textbf{3FGA} & \textbf{FGA} \\
		\hline
		Men D1       & 25.41 & 12.70 & 12.24 & 2.73 & 7.58 & 5.53 & 4.19 & 4.92  & 1.40  & 0.42  & -0.87 \\
		Men D2       & 23.58 & 13.83 & 5.67  & 3.92 & 3.93 & 3.53 & 3.38 & 3.42  & 2.78  & 1.42  & -0.17 \\
		Men D3       & 17.33 & 12.61 & 3.40  & 5.76 & 1.89 & 3.57 & 3.98 & 2.02  & 1.28  & 0.45  & 0.49  \\
		Women D1       & 14.13 & 11.35 & 7.55  & 5.73 & 5.94 & 5.13 & 4.99 & 2.63  & 2.70  & 1.86  & 0.65  \\
		Women D2       & 13.17 & 12.08 & 5.58  & 3.57 & 4.37 & 3.80 & 3.54 & 2.51  & 1.97  & 0.01  & 0.62  \\
		Women D3       & 11.86 & 12.45 & 3.39  & 6.86 & 2.23 & 4.09 & 5.12 & 1.61  & 3.61  & 1.86  & 1.17  \\
		\hline\noalign{\smallskip}
		\multicolumn{12}{p{17cm}}{\raggedright Values are estimated using Poisson regression. Columns are sorted in order of mean home advantage. Note that all values represent improvements (decrease in PF and TOV and increase in all other statistics).}
	\end{tabular}
\end{table}

Finally, we can use the $\textrm{RPI}'$ coefficients to examine which statistics are most impacted by team strength. As expected, with only one exception (3FGA for men's D2) stronger teams in all gender-division combinations have superior performance in all examined statistics. Interestingly, the $\textrm{RPI}'$ coefficients are highly correlated (0.76) with the home advantage coefficients. Thus, it may be the case that scorekeeper biases are also correlated with strength and stronger teams (and their players) record improved box score statistics not only due to their strength, but also due to their scorekeepers.

\section{Conclusions and Discussion}
\label{conclusions_and_discussion}
In this paper we have presented an examination of the impact of home court on box score statistics in NCAA basketball. Teams playing at home received a boost in nearly all statistical categories across all gender-division combinations and these results have remained consistent over five seasons of data. We also made use of neutral location games to separate the impact of being the home team from the impact of being the away team, providing a unique view into home court advantage. 

This paper has presented substantial evidence supporting the idea of scorekeeper and referee biases impacting box score statistics. In both cases, external factors seem to influence these decision makers. For scorekeepers, gender and division appear to impact their behavior, since men and teams in higher divisions receive a greater boost in most statistics at home compared to women and teams in lower divisions. While these biases may be unintentional, they can inflate the perceived ability of men and higher divisions, leading to greater attention for those leagues at the expense of women and lower divisions. Scorekeepers may also be influenced by the strength of the team that hired them, as the impacts of home court and team strength are highly correlated for box score statistics. For referees, we presented strong evidence that larger crowds have greater impacts on their decisions, specifically increasing their bias in favor of the home team. Through foul trouble and free throws, such biases have direct impacts on the games, and can steal wins from better performing teams, simply because they are playing on the road. 

There are several ideas presented in this paper which we believe are deserving of further analysis. The concept of scorekeeper and referee biases could be further evaluated using structural equation models or other instrumental variable methods. Such approaches have the potential to provide key insights into the relationships to other factors not examined in this work. Additionally, a logical extension of the work presented here would be to develop robust adjustment methods for correcting the biases and inconsistencies present in NCAA basketball. Such methods would allow counterfactual considerations, such as evaluating what would have likely happened if a game had been played in a different location. Careful attention would need to be paid to both the statistical modeling techniques underlying such adjustment methods and to the basketball related assumptions and implications inherent in any adjustments. However, given the obvious difficulty in removing human biases from the game, a well designed adjustment process would go a long way in allowing  box score statistics and game results to truly reflect the on-court performance of teams and players. 

\bibliographystyle{apalike}
\bibliography{ncaa_paper_refs} 

\end{document}